\documentclass[aps,prl,preprint,superscriptaddress]{revtex4}
\usepackage{graphicx}

\begin{document}

\begin{flushright}
  March, 2003 \ \ \\
  OU-HEP-433 \ \\
\end{flushright}


\title{The nonperturbative origin of delta-function singularity\\
in the chirally-odd twist-3 distribution function $e(x)$}


\author{M.~Wakamatsu and Y.~Ohnishi}
\affiliation{Department of Physics, Faculty of Science, \\
Osaka University, \\
Toyonaka, Osaka 560, JAPAN}



\begin{abstract}
We analytically prove that the existence of the delta-function
singularity in the chirally-odd twist-3 distribution $e(x)$ of the
nucleon is inseparably connected with the nonvanishing quark
condensate as a signal of the spontaneous chiral symmetry
breaking of the QCD vacuum. This singularity in $e(x)$, which would be
observed as a sizable violation of the 1st moment sum rule, is then
interpreted as giving a very rare case that the nontrivial vacuum
structure of QCD manifests in an observable of a localized QCD
excitation, i.e. the nucleon.
\end{abstract}


\maketitle


In spite of several theoretically interesting properties,
the chirally-odd twist-3 distribution $e(x)$ has been thought
of as an academic object of study, since it is chirally odd
and will not appear in deep-inelastic scatterings if quark masses
are ignored \cite{JJ92}.
Very recently, however, the CLAS collaboration gave the first
information on this interesting quantity through the measurement of
the azimuthal asymmetry $A_{L U}$ in the electro-production of pions
from deeply inelastic scatterings of longitudinally polarized electrons
off unpolarized protons \cite{CLAS02}. (See also \cite{HERMES00},
\cite{HERMES01}.)
Why is this distribution function so interesting?
First of all, its first moment is related to the well-known
$\pi N$ $\Sigma$-term, which is believed to give important 
information on the explicit chiral symmetry breaking of
QCD \cite{JJ92}.
Despite the fact that the Regge argument throws some doubts on 
the convergence of this first-moment sum rule, it has
been expected to give some direct information on the scalar 
charge and/or the scalar density of the nucleon, which can otherwise
be obtained only indirectly through the analysis of low energy
pion-nucleon scattering amplitude \cite{JLS73}.
Very interestingly, within the framework of perturbative QCD, 
Burkardt and Koike found the existence of delta-function singularity 
at the Bjorken variable $x = 0$ in $e(x)$ \cite{BK02}.
They showed that in order 
to satisfy the $\Sigma$-term sum rule one needs to introduce 
$\delta(x)$ term. Since the point $x = 0$ is experimentally inaccessible,
this in turn indicates that experimental measurement can never confirm 
the $\Sigma$-term sum rule but would rather confirm the violation of it.
What is the physical origin of $\delta(x)$ term in $e(x)$, then?
This question has been addressed by Efremov et al. in their recent
studies \cite{ES03}-\nocite{EGS01}-\cite{EGS02}.
They conjectured that the origin of $\delta(x)$ singularity
in $e(x)$ can be traced back to the nonvanishing vacuum quark condensate
of QCD vacuum. This is quite reasonable if one remembers the inseparable
connection between the nucleon scalar charge and the vacuum quark condensate.
Unfortunately, no explicit proof is given in their paper.
The purpose of the present paper is to give a complete analytical
proof of the fact that the $\delta(x)$ singularity in $e(x)$ is in fact
originates from the nonvanishing quark condensate of the QCD
vacuum. The explicit proof is given within the framework of the chiral
quark soliton model (CQSM). However, it will be sure that
the proved fact itself holds in more general ground, since what is
essential here is the realization in nature of the nonvanishing vacuum
quark condensate of QCD.

We start with the general definition of the distribution function $e(x)$
given as :
\begin{equation}
 e(x) = M_N \int_{-\infty}^{\infty} \frac{d z_0}{2 \pi} \,\,
 e^{-i x M_N z_0} 
 \left. \langle N | \bar{\psi} (0) \psi (z) | N \rangle 
 \right|_{z_3 = -z_0, z_\perp = 0} ,
\end{equation}
in the quark-only approximation \cite{DPPPW96},\cite{DPPPW97}.
Here, we concentrate on the isoscalar piece 
$e^{(I = 0)} (x) = e^u (x) + e^d (x)$ of $e(x)$, simply expressing it
as $e(x)$. Within the framework of the CQSM,
the leading contribution (in the $1/N_c$ expansion) to 
this distribution is given in the following form \cite{DPPPW96}
\nocite{DPPPW97}\nocite{WK98}--\cite{WK99} :
\begin{equation}
 e(x) = M_N \int \frac{d z_0}{2 \pi} \,e^{-i x M_N z_0} \,E_U (z_0) ,
\end{equation}
with
\begin{equation}
 E_U (z_0) = N_c \int \frac{d^3 p}{(2 \pi)^3} \,\sum_{n \leq 0} \,
 e^{i E_n z_0} \,\Phi_n^{\dagger} (\mbox{\boldmath $p$}) \,\gamma^0
 e^{i p_3 z_0} \,\Phi_n (\mbox{\boldmath $p$}) .
\end{equation}
Here $E_n$ and $\Phi_n$ are the eigenenergies and the corresponding 
eigenfunctions of the Dirac Hamiltonian $H$ with the static soliton 
background, i.e.
\begin{equation}
 H \Phi_n = E_n \Phi_n , \ \ \ \ \ 
 H = \frac{\mbox{\boldmath $\alpha$} \cdot \nabla}{i} + M \beta \,
 e^{\,i \gamma_5 \mbox{\boldmath $\tau$} \cdot 
 \hat{\mbox{\boldmath $r$}} F(r)} .
\end{equation}
The summation $\sum_{n \leq 0}$ is meant to be taken over the so-called 
valence-quark orbital (it is the lowest energy eigenstate that emerges
from the positive-energy Dirac continuuum) as well as all the
negative-energy Dirac-sea orbitals. Actually, we are interested in the
nucleon observables measured in reference 
to the physical vacuum, so that $E_U (z_0)$ should be replaced by 
\begin{equation}
 E_U (z_0) \rightarrow E (z_0) \equiv E_U (z_0) - E_{U = 1} (z_0) .
\end{equation}
Here the vacuum subtraction term $E_{U = 1} (z_0)$, which will be denoted
as $E_{vac}(z_0)$ hereafter, is obtained from $E_U (z_0)$ by setting 
$U = 1$ or $F (r) = 0$, and by  excluding the sum over the valence level.
To be more explicit, it is given as 
\begin{equation}
 E_{vac} (z_0) = N_c \int \frac{d^3 p}{(2 \pi)^3} \sum_{k \,(E_k^{(0)} < 0)}
 e^{i E_k^{(0)} z_0} \,\Phi_k^{(0) \dagger} 
(\mbox{\boldmath $p$}) \,\gamma^0 \,
 e^{i p_3 z_0} \,\Phi_k^{(0)} (\mbox{\boldmath $p$}) ,
\end{equation}
with
\begin{equation}
 H_0 \Phi_k^{(0)} = E_k^{(0)} \Phi_k^{(0)}, \ \ \ \ \ 
 H_0 = \frac{\mbox{\boldmath $\alpha$} \cdot \nabla}{i} + M \beta .
\end{equation}
Now, a crucial observation is that the existence of $\delta(x)$ singularity 
in $e(x)$ would mean that $E(z_0)$ contains a piece that survives even
in the $z_0 \rightarrow \infty$ limit, i.e. the existence of infinite range
quark-quark correlation. To understand that this behavior of $E(z_0)$ is 
essentially due to the nonvanishing quark condensate of QCD vacuum, it is
instructive to first investigate the vacuum term $E_{vac} (z_0)$.
Since $\Phi_k^{(0)}$ is basically the plane-wave quark states with the 
effective mass $M$, we easily find that 
\begin{equation}
 E_{vac} (z_0) = - 4 N_c M \,V \,\int \frac{d^3 k}{(2 \pi)^3} \,
 \frac{e^{- i \,\sqrt{k^2 + M^2} \,z_0}}{\sqrt{k^2 + M^2}} \,e^{i k_3 z_0} .
 \label{lcone}
\end{equation}
Here $V$ is the total volume of the system, which is of course infinite.
However, we can later show that the difference $E(z_0) = E_U (z_0) 
- E_{vac} (z_0)$ is finite, so that let us continue the argument.
Here, we notice an important relation : 
\begin{equation}
 E_{vac} (z_0) = - 4 N_c M \,V \cdot \left. \left\{ \Delta^{(1)} (z ; M^2) 
 + i \,\Delta (z ; M^2) \right\}  \right|_{z_3 = -z_0, z_\perp = 0},
\end{equation}
where $\Delta^{(1)}$ and $\Delta$ defined by 
\begin{eqnarray}
 \Delta^{(1)} (z ; M^2) &=& \int \frac{d^3 p}{(2 \pi)^3} \,\,
 \delta (p^2 - M^2) \,e^{-i p z} , \\
 \Delta (z ; M^2) &=& \int \frac{d^3 p}{(2 \pi)^3 \,i} \,\,
 \epsilon (p_0) \,\delta (p^2 - M^2) \,e^{-i p z} ,
\end{eqnarray}
with $\epsilon (u) = u / |u|, \epsilon (0) = 0$, are the invariant delta 
(Green's) functions found in standard textbooks \cite{R68}.
What is meant by Eq.(\ref{lcone}) is that
$E_{vac} (z_0)$ is given as the  light-cone limit of the two invariant 
delta functions. (Note that $z^2 \equiv z_0^2 - (z_3^2 + z_1^2) 
\rightarrow 0$ as $z_3 \rightarrow - z_0, z_1 \rightarrow 0$.)
We need here the behavior of $\Delta^{(1)} (z ; M^2)$ and $\Delta (z ; M^2)$ 
near the light-cone $z^2 = 0$, which is given as 
\begin{eqnarray}
 \Delta^{(1)} (z : M^2) \! &=& \! - \frac{1}{2 \pi^2} \,
 \left\{ P \frac{1}{z^2}
 - \frac{M^2}{4} \,\ln \left(M^2 | z^2 | \right)
 \ + \ \frac{M^2}{2} \, \left(\ln 2 - \gamma + \frac{1}{2} \right)
 + O (z^2) \right\} , \ \ \\
 \Delta(z ; M^2) \! &=& \! - \frac{1}{2 \pi} \,\epsilon (z_0) \,
 \left\{ \delta (z^2) - \frac{M^2}{4} \,\theta (z^2) + O (z^2) \right\} .
\end{eqnarray} 
One sees that these contain several terms that diverge on the light-cone.
Note, however, that the CQSM is defined with some approriate regularization.
In fact, without regularization, the vacuum quark condensate is quadratically
divergent. It was shown in \cite{KWW99} that the Pauli-Villars
regularization scheme with two subtraction terms eliminates all the
divergences including the most singular vacuum quark condensate.
In our present problem, the double-subtraction Pauli-Villars scheme means 
the replacement of $E_{vac} (z_0)$ by $E_{vac}^{reg} (z_0)$ defined as 
\begin{equation}
 E_{vac}^{reg} (z_0) \equiv E_{vac} (z_0) - \sum_{i = 1}^2 \,c_i
 \left(\frac{\Lambda_i}{M} \right) \,E_{vac}^{\Lambda_i} (z_0) .
\end{equation}
Here $E_{vac}^{\Lambda_i} (z_0)$ is obtained from $E_{vac} (z_0)$ by replacing
the mass parameter $M$ by $\Lambda_i$.
It was shown in \cite{KWW99} that, if the parameters $c_1, c_2,
\Lambda_1$, and $\Lambda_2$ are chosen to satisfy the two conditions : 
\begin{eqnarray}
 1 - \sum_{i = 1}^2 \,c_i \,\left(\frac{\Lambda_i}{M} \right)^2 &=& 0 , 
 \label{cond1} \\
 1 - \sum_{i = 1}^2 \,c_i \,\left(\frac{\Lambda_i}{M} \right)^4 &=& 0 . 
 \label{cond2}
\end{eqnarray}
the quadratic as well as logarithmic divergences in the vacuum quark
condensate are completely eliminated.
Now it is easy to verify that this regularization works
perfectly also for $E_{vac} (z_0)$. The condition (\ref{cond1})
eliminates the quark mass independent terms including $\delta(z^2)$ and $P
\frac{1}{z^2}$, while the condition (\ref{cond2}) removes the terms
proportional to $M^2$ including $M^2  \theta (z^2)$ and $M^2 \ln |z^2|$. 
With the understanding that the light-cone limit $z_3 \rightarrow -z_0, 
z_{\perp} \rightarrow 0$ is taken after this Pauli-Villars subtraction, 
we are therefore left with
\begin{equation}
 E_{vac}^{reg} (z_0) = - \frac{N_c M}{2 \pi^2} \,\left\{ M^2 \,\ln M^2
 - \sum_{i = 1}^2 \,c_i \left(\frac{\Lambda_i}{M} \right)^2
 \Lambda_i^2 \ln \Lambda_i^2 \right\} .
\end{equation}
Very interestingly, the r.h.s. of the above equation is actually independent 
on $z_0$ and it perfectly coincides with the expression of the vacuum quark 
condensate given in \cite{KWW99}, i.e. we find that
\begin{equation}
 E_{vac}^{reg} (z_0) = V \,\langle \bar{\psi} \psi \rangle_{vac} .
\end{equation}
We have so far confirmed that the non-zero vacuum quark condensate can in fact
generate a constant term in $E (z_0)$ and consequently a $\delta (x)$ 
singularity in $e(x)$. The proof is not yet completed, since the vacuum
contribution $E_{vac}^{reg} (z_0)$ is actually proportional to an infinite 
volume $V$. Note however that the physical distribution $e(x)$ of the
nucleon is the Fourier transform of the difference function 
$E_U (z_0) - E_{vac} (z_0)$. The exact manipulation of the first term 
$E_U (z_0)$ requires numerical analysis, but for our present purpose, an
approximate treatment with use of the derivative expansion is enough and
even more instructive. To this end, we start with the Green's function 
representation of the distribution function derived by Diakonov et al.
\cite{DPPPW96} as
\begin{equation}
 e(x) = e_2 (x) + e_2^* (x) ,
\end{equation}
with
\begin{eqnarray}
 e_2 (x) &=& N_c \,M_N \,\int_0^{\infty} \,
 \frac{d z_0}{2 \pi} \,e^{-i x M_N z_0} \nonumber \\
 &\times& \!\! \int d^3 \mbox{\boldmath $R$} \,
 \left\{ \mbox{Tr} \,i \,
 G (-z_0, z_0 \mbox{\boldmath $e$}_3 - \mbox{\boldmath $R$} \,; 
 0 , -\mbox{\boldmath $R$})
 - \mbox{Tr} \,i \,G^{(0)} (-z_0, z_0 \mbox{\boldmath $e$}_3 
 - \mbox{\boldmath $R$} \,; 0, -\mbox{\boldmath $R$}) \right\} . \ \ 
 \label{greenrep}
\end{eqnarray}
Here $G$ and $G^{(0)}$ are single quark Green's functions with and without 
the soliton background : 
\begin{eqnarray}
 G (x^0, \mbox{\boldmath $x$} \,;\, y^0 \mbox{\boldmath $y$}) 
 \ &=& - \,
 \left\langle x^0, \mbox{\boldmath $x$}
 \left| \frac{1}{i \not\!\partial - M U^{\gamma_5}} \right|
 y^0, \mbox{\boldmath $y$} \right\rangle , \\
 G^{(0)} (x^0, \mbox{\boldmath $x$} \,;\, y^0, \mbox{\boldmath $y$})
 &=& - \,\,
 \left\langle x^0, \mbox{\boldmath $x$}
 \left| \frac{1}{i \not\!\partial - M} \right| 
 y^0 \mbox{\boldmath $y$} \right\rangle .
\end{eqnarray}
Using the standard technique of derivative expansion, the leading contribution 
becomes
\begin{eqnarray}
 &\,& G (x^0, \mbox{\boldmath $x$} \,;\, y^0 \mbox{\boldmath $y$}) 
 - G^{(0)} (x^0, \mbox{\boldmath $x$} \,;\, y^0 \mbox{\boldmath $y$})
 \nonumber \\
 &=& M \,\left[ U^{\gamma_5} (\mbox{\boldmath $y$}) - 1 \right] \,
 \int \frac{d^4 k}{(2 \pi)^4} \,
 e^{-i k \cdot (x - y)} \,\frac{1}{M^2 - k^2 - i \varepsilon} 
 \ + \ \cdots , \label{deriv}
\end{eqnarray}
where the indicated higher derivative terms will be neglected hereafter.
Now putting (\ref{deriv}) into (\ref{greenrep}), we finally obtain
\begin{equation}
 e(x) = M_N \,\int_{-\infty}^{\infty} \,d z_0 \,e^{-i x M_N z_0} \,
 E(z_0) , \label{esmall}
\end{equation}
with
\begin{eqnarray}  
 E(z_0) &=& \frac{1}{8} \,\int d^3 \mbox{\boldmath $R$} \,\,
 \mbox{Tr} \,\left[U^{\gamma_5} (\mbox{\boldmath $R$}) 
 - 1 \right] \nonumber \\
 &\,& \ \times \  (-4 N_c M) \,\left. 
 \left\{ \Delta^{(1)} (z ; M^2) + i \Delta (z ; M^2) \right\} 
 \right|_{z_3 = -z_0, z_{\perp} = 0} .
\end{eqnarray}
It is now evident that, after regularization, this gives
\begin{equation}
 E (z_0) = \int d^3 r \,[\cos F (r) - 1] \times
 \langle \bar{\psi} \psi \rangle_{vac} .
\end{equation}
To rewrite it further, we recall the relation for the $\pi N$ $\Sigma$-term 
and/or the nucleon scalar charge $\bar{\sigma}$ \cite{KWW99},\cite{W92},
\begin{equation}
 \Sigma_{\pi N} = m_0 \,\bar{\sigma} = 
 - f_{\pi}^2 \,m_{\pi}^2 \,\int \,d^3 r \,[\cos F (r) - 1 ] ,
\end{equation}
together with the familiar Gell-Mann-Oakes-Renner relation
\begin{equation}
 m_0 \,\langle \bar{\psi} \psi \rangle_{vac} = - f_{\pi}^2 \,m_{\pi}^2 .
\end{equation}
These altogether give
\begin{equation}
 E (z_0) = \bar{\sigma} . \label{elarge}
\end{equation}
We thus find that $E (z_0)$ is $z_0$ independent constant as
$E_{vac}^{reg} (z_0)$, but now it is a finite one, representing the nucleon
scalar charge measured with respect to the physical vacuum having uniform
scalar charge density of negative sign.

\begin{figure}[htb] \centering
\begin{center}
 \includegraphics[width=12.0cm]{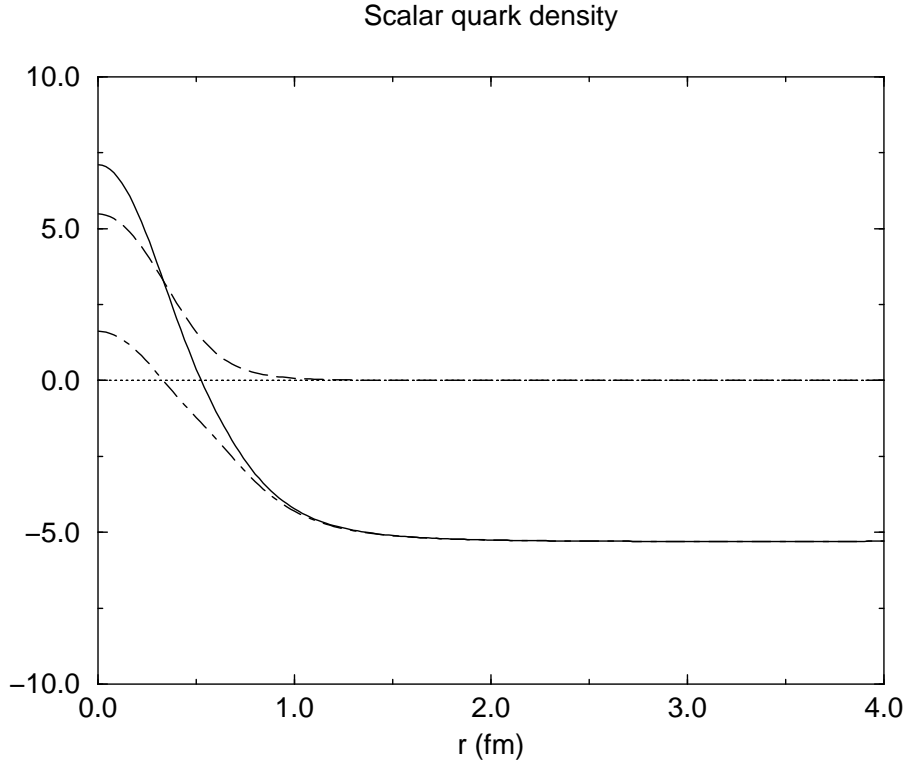}
\end{center}
\vspace*{-0.5cm}
\renewcommand{\baselinestretch}{1.20}
\caption{The scalar quark density around the nucleon center
predicted by the CQSM.
The dashed and dash-dotted curves are respectively the contribution
of $N_c$ valence quarks and that of the negative energy
Dirac-sea quarks in the soliton background, while the solid curve
is their sum. \label{sdens}}
\end{figure}

The deep reason why $E (z_0)$ contains a constant term which does not
attenuate even in the $z_0 \rightarrow \infty$ limit can be understood
as follows. The solid curve in Fig.\ref{sdens} represents the scalar
quark density predicted by the CQSM.
It is actually a sum of the two terms : the one (represented by
the dashed curve) is the contribution of $N_c$ valence quarks, and
the other (illustrated by the dash-dotted curve) is that of the 
negative-energy Dirac-sea quarks in the soliton background.
While the valence quark contribution rapidly damps as the distance $r$
from the soliton center becomes large, the contribution of the Dirac-sea
quarks approaches a nonzero (negative) constant which is nothing but the
vacuum quark condensate \cite{W92}.
Consequently, the nucleon scalar charge is not
a naive spatial integral of this density, but it is a difference from the
corresponding scalar charge of the physical vacuum, which itself is
of course infinite, i.e.
\begin{eqnarray}
 \bar{\sigma} &=& \int d^3 x \,
 \left[ \langle N | \bar{\psi} \psi (x) | N \rangle
 - \langle \bar{\psi} \psi \rangle_{vac} \right]
 \ = \ \int d^3 x \,\langle N | \bar{\psi} \psi (x) | N \rangle
 - V \,\langle \bar{\psi} \psi \rangle_{vac} .
\end{eqnarray}
Similarly, both of $E_U (z_0)$ and $E_{vac} (z_0)$ contain constant terms 
with infinite magnitude but their difference is a finite constant, which 
is nothing but the nucleon scalar charge $\bar{\sigma}$. 
Now, from (\ref{elarge}) and (\ref{esmall}), it follows that
\begin{equation}
 e (x) = \bar{\sigma} \,\delta (x) ,
\end{equation}
and the nonperturbative origin of the delta-function singularity in $e (x)$ 
is clarified. This results is somewhat misleading, however. While it certainly 
satisfies the 1st moment sum rule, 
\begin{equation}
 \int_{-1}^1 \,e (x) \,d x \ = \ \bar{\sigma} \ = \ 
 \frac{\sum_{\pi N}}{m_0} , \label{sumrule}
\end{equation}
the literal acceptance of the result (\ref{sumrule}) would mean that $e (x)$
is a trivial function which is zero everywhere aside from the delta-function
singularity at $x = 0$. 
Undoubtedly, this physically implausible conclusion comes from an approximate 
treatment based on the derivative expansion.
As is well-known, the derivative 
expansion is an asymptotic expansion, and a naive application of it sometimes
leads to inconsistencies. For instance, the famous topological baryon number
current is known to emerge from the third derivative term of the expansion,
but the simultaneous account of it with the explicit contribution of the $N_c$
valence quarks would lead to a double counting.
The exact calculation without recourse to the derivative expansion type 
approximation is possible only numerically. Our preliminary calculation shows
that the contribution of the $N_c$ valence quarks to $e (x)$ in fact has 
nonzero distribution peaked around $x \simeq 1/3$. (Sea also. \cite{ES03}.)
It also indicates that the contribution of the Dirac-sea quarks to $e (x)$ 
has nonzero support also for $x \neq 0$ in addition to the delta-function 
singularity at $x = 0$. The results of this numerical calculation for $e (x)$
within the framework of the CQSM will be reported elsewhere.

In summary, we have clarified the physical (nonperturbative) origin of the 
delta-function singularity in the chirally-odd twist-3 distribution function 
$e (x)$ of the nucleon. It can be traced back to the long-distance quark-quark
correlation of scalar type, which signals the spontaneous chiral symmetry 
breaking of the QCD vacuum. Although the explicit proof has been given
within the framework of the CQSM, it is evident that the proved fact itself
holds in more general ground. What is essential in the proof is the
existence of the nontrivial long-range quark-quark correlation of scalar
type as illustrated in Fig.1 for the quark scalar density around the nucleon
center. One might suspect if such a delta-function singularity in a
distribution function is merely an academic object, since $x=0$ is
experimentally inaccessible.
However, its existence can in principle be observed as a
sizable violation of the $\Sigma$-term sum rule. If it is in fact confirmed,
it would provide us with an unprecedented example in which the nontrivial
structure of QCD vacuum manifests in the structure of a localized QCD
excitation, i.e. the nucleon.
A natural question is whether the delta-function singularity of similar
nature is expected also in other distribution functions of the nucleon.
We think it unlikely, since only the scalar type quark-quark correlation
can have nonzero expectation value in the uniform vacuum.
More careful analysis may be necessary, however, because the similar 
delta-function singularity is indicated also for the twist-3 distribution 
$h_L (x)$ on the ground of perturbative QCD \cite{BK02}.

\begin{acknowledgments}
This work is supported in part by a Grant-in-Aid for Scientific
Research for Ministry of Education, Culture, Sports, Science
and Technology, Japan (No.~C-12640267)
\end{acknowledgments}

\bibliographystyle{unsrt}
\bibliography{edelta}

\end{document}